# Predicting the Activity and Selectivity of Bimetallic Metal Catalysts for Ethanol Reforming using Machine Learning

Nongnuch Artrith[1,*], Zhexi Lin[1], Jingguang G. Chen[1,2,*]

[1]*Department of Chemical Engineering, Columbia University, New York, NY, USA.*
[2]*Chemistry Division, Brookhaven National Laboratory, Upton, NY, USA.*
Corresponding authors: *nartrith@atomistic.net *jgchen@columbia.edu

**Abstract:** Machine learning is ideally suited for the pattern detection in large uniform datasets, but consistent experimental datasets on catalyst studies are often small. Here we demonstrate how a combination of machine learning and first-principles calculations can be used to extract knowledge from a relatively small set of experimental data. The approach is based on combining a complex machine-learning model trained on a computational library of transition-state energies with simple linear regression models of experimental catalytic activities and selectivities from the literature. Using the combined model, we identify the key C–C bond scission reactions involved in ethanol reforming and perform a computational screening for ethanol reforming on monolayer bimetallic catalysts with architectures TM-Pt-Pt(111) and Pt-TM-Pt(111) (TM = 3d transition metals). The model also predicts four promising catalyst compositions for future experimental studies. The approach is not limited to ethanol reforming but is of general use for the interpretation of experimental observations as well as for the computational discovery of catalytic materials.

Ethanol reforming to carbon monoxide and hydrogen is an attractive means for hydrogen production,[1–3] but catalytic ethanol decomposition can also lead to undesired methane production or to total decomposition into atomic carbon and oxygen that lead to the deactivation of the catalyst.[4] Bimetallic catalysts based on platinum (Pt) have previously been identified as promising catalysts for ethanol reforming with good selectivity for the reforming reaction,[5] but only few compositions have been extensively characterized experimentally.

Skoplyak *et al.* reported a quantification of the ethanol decomposition activity and selectivity of Pt and six Pt-based core-shell surfaces using temperature-programmed desorption (TPD) spectroscopy measurements.[4,6,7] Architectures both with a monolayer of a transition metal (TM) on top of the Pt(111) surface, TM-Pt-Pt(111), and with a subsurface TM monolayer, Pt-TM-Pt(111), were investigated for TM = Ti, Fe, and Ni.

An exhaustive experimental investigation of additional core-shell compositions for ethanol reforming would be time consuming and would quickly become infeasible when compositions with more than two metal species are considered or when Pt is replaced with earth-abundant TM carbides.[8,9] Instead, an understanding of the atomic-scale factors that control the catalytic activity and selectivity would be preferable, as it would potentially enable the rational design of improved catalysts.

Machine learning (ML) has previously been used to accelerate first-principles calculations of catalyst materials,[10–18] but incorporating experimental data is challenging. In this letter, we demonstrate how a combination of ML and first-principles calculations can be employed to interpret experimental activity and selectivity data and to predict the catalytic performance of additional bimetallic catalysts for ethanol reforming. Since experimental activity and selectivity data, obtained with consistent methods, are relatively scarce, we developed a two-step approach that augments the small experimental data sets with extensive computational data. The approach is motivated by the assumption that the activity and selectivity for ethanol reforming are determined by the kinetically relevant reactions of all competing ethanol decomposition reactions. This means, knowledge of the reaction energies and the kinetic activation energies of the different reaction pathways should be sufficient to predict both the activity and selectivity.

Thermochemical reaction profiles can be efficiently estimated from automated first-principles density-functional theory (DFT) calculations.[19,20] For the present work, we performed DFT calculations using the *Vienna Ab Initio Simulation Package* (VASP)[21,22] and all input files were automatically generated using the *pymatgen* toolkit[23] (see **Section S1** in the **Supporting Information** for further details of our computational approach). The activation energies of intermediate reaction steps can, in principle, also be calculated with DFT, for example, using the nudged-elastic band (NEB) method.[24–26] However, NEB calculations are computationally demanding and not well suited for automated high-throughput calculations. In the present work, we adopted a recently introduced machine-learning accelerated NEB method by Garrido Torres *et al.*[27–29] as implemented in the atomic simulation environment (ASE) package,[30,31] which yields DFT activation energies at a reduced computational cost. But owing to the large number of possible reaction pathways for ethanol decomposition, calculating all activation energies for a large number of different catalyst compositions would nevertheless be a formidable challenge.

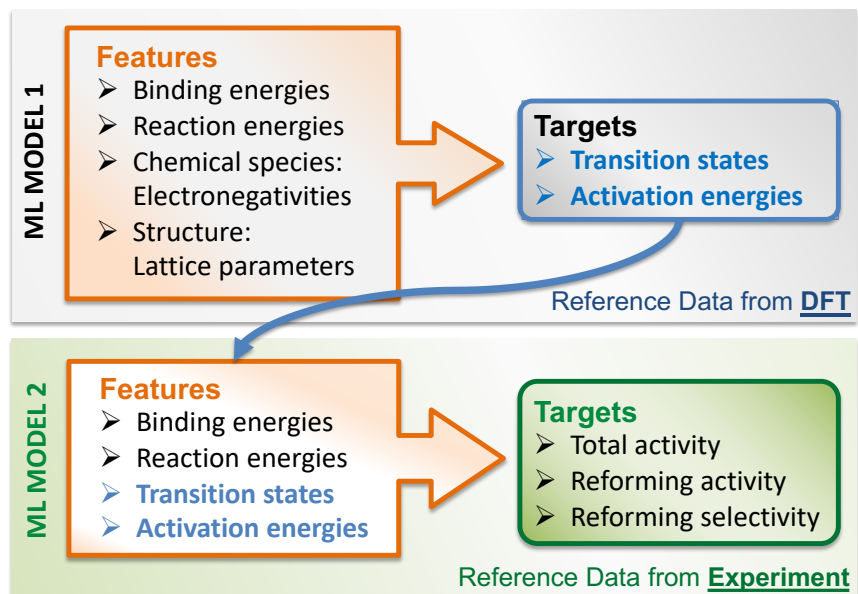

**Figure 1** Flowchart of the combined machine-learning (ML) approach consisting of two ML models. ML Model 1 is a non-linear model combining Random Forest Regression (RFR) and Gaussian Process Regression (GPR) trained on extensive reference data from density-functional theory (DFT) calculations to predict transition-state energies of ethanol decomposition reaction steps. The predicted transition-state energies enter a second linear model (ML Model 2) that is trained on a smaller data set of catalytic activities and selectivities from published experiments.

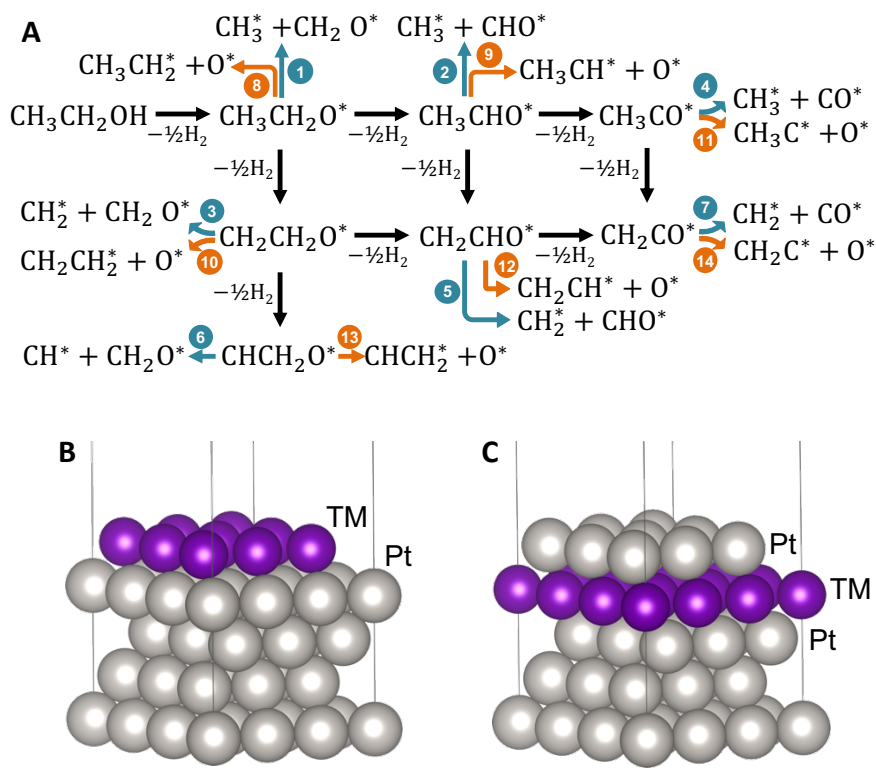

**Figure 2** **(A)** Network of reaction pathways leading to C–C and C–O scission during the ethanol decomposition reaction. A star (*) indicates that the species is adsorbed on the catalyst surface. C–C and C–O scission reactions are highlighted in blue and orange, respectively. **(B)**, **(C)** Core-shell surface structure models with **(B)** a monolayer of transition-metal species TM on top of the Pt(111) surface, TM-Pt-Pt(111), and **(C)** a TM monolayer below the top atomic layer of Pt atoms on the Pt(111) surface, Pt-TM-Pt(111).

Our approach to circumvent the need for extensive NEB calculations is, first, to construct a ML model for the prediction of transition-state energies from the thermochemical reaction energies (Model 1). Then, a second ML model is trained to predict the catalytic activity and selectivity based on all transition-state energies (Model 2). Both models together allow the prediction of catalytic activities/selectivities directly from principal chemical properties and features that can be efficiently determined with high-throughput DFT calculations. A flowchart of the combined approach is shown in **Figure 1**.

The transition-state ML Model 1 was trained on a computational database containing the relative DFT energies of the 14 different ethanol decomposition pathways shown in the reaction network in **Figure 2**. Only reaction paths until the first C–C or C–O bond scission were evaluated, since the kinetic barrier associated with these steps can be assumed to be rate determining for the entire decomposition reaction.[32] In addition to the thermochemical reaction energies, the kinetic activation energies for the C–C and C–O bond scissions were also calculated (highlighted in blue and orange in **Figure 2**). The calculations were performed over the pure Pt(111) surface and the monolayer core-shell architectures TM-Pt-Pt(111) and Pt-TM-Pt(111) with TM = Ni, Cu. Representations of the periodic slab structure models with a 3×3 surface unit cell are shown in **Figure 2 (B)** and **(C)**, and further details of the structure models are given in **Section S2** of the **Supporting Information**. The final reference data set comprised a total of 78 activation energies from our own calculations. Additionally, we included reaction energies and transition-state energies from 41 ethanol decomposition reactions over Pd and PdAu alloys taken from the literature,[32] giving a total number of 119 data points (**Table S1**).

We chose two different ML techniques, (i) random forest regression (RFR)[33,34] and (ii) Gaussian process regression (GPR),[35] to train on the DFT transition-state energies (the *targets*) within this dataset. Model inputs (i.e., the *features*) were the reaction energies of the C–C and C–O scission reactions, as well as features that distinguish between the chemical species (electronegativity[36] and nearest-neighbor distance[37]) and the surface reactivity (DFT adsorption energy of ethanol). All details of the model construction are given in **Section S3** of the **Supporting Information**.

The best models obtained with both ML techniques achieve an individual root-mean-squared error (RMSE) of 0.35 eV for the transition state energies (**Figures S1**), as determined by leave-one-out *cross validation* (CV). The RMSE of an ensemble model combining RFR and GPR was 0.31 eV with a mean absolute error (MAE) of 0.20 eV. The ensemble model was therefore used in the subsequent calculations.

We note that it is a common approximation to assume that activation energies are linear functions of the reaction energies (Brønsted-Evans-Polanyi relationship[38,39]). As shown in supporting **Figure S2**, this principle is valid for the present systems but yields a mean absolute error that is 75% greater (0.35 eV) and a root mean squared error that is 45% greater (0.45 eV) than the corresponding errors of ML model I.

A comparison of the predicted transition-state energies with the DFT reference values is shown in **Figure 3**. Note that the energy of the transition states varies over a range of 5 eV, so that an accuracy of ~0.3 eV corresponds to an uncertainty of ~6%. Based on a convergence analysis of our transition-state calculations and on previous discussion in the literature,[40] we expect that the intrinsic error of the NEB energies is close to 0.1 eV. Remarkably, ML Model 1 predicts the transition-state energies of the core-shell catalysts and those of the alloy compositions equally well.

For additional model validation, we used ML Model 1 to predict the transition-state energies for reactions over Pt-Fe-Pt(111), which is highly selective for ethanol reforming and more active than Pt(111).[7] The results are compared with actual DFT calculations in **Figure S3**. As seen in the figure, the ML model is in good agreement with DFT results, and the RMSE of the predictions (0.22 eV) is in line with our expectations based on the CV score of the ML model (0.31 eV).

Using ML Model 1, we predicted the transition-state energies for the ethanol decomposition reactions over the remaining core-shell catalysts for which experimental activity and selectivity data is available: Ti-Pt-Pt(111), Pt-Ti-Pt(111), Fe-Pt-Pt(111), and Pt-Fe-Pt(111). Along with Pt(111) and the Ni-Pt core-shell catalysts, these compositions form our experimental reference dataset, comprising a total of 7 data points.[7] Intuitively, we expect that out of all possible reaction pathways there are only few that are independent and contribute to the overall ethanol decomposition. If this hypothesis is correct, 7 data points might be sufficient for the construction of a simple predictive model for obtaining reforming activity and selectivity.

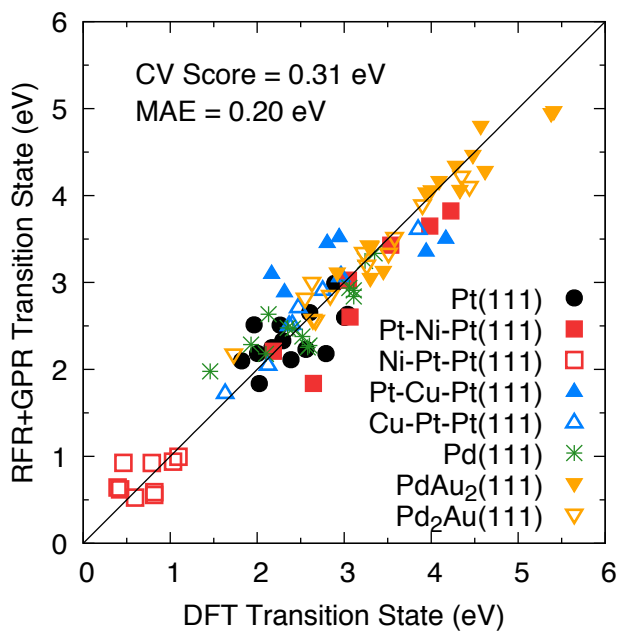

**Figure 3** Transition state energy predicted by the RFR+GPR ML Model 1 versus the DFT reference energies. The plotted points were not included in the model construction and are predictions based on leave-one-out cross validation (CV). The root-mean squared error as estimated via CV is 0.31 eV, and the mean absolute error (MAE) is 0.20 eV.

The activity and selectivity quantification performed by Skoplyak et al.[7] assumed that ethanol decomposition involves the following three principal reactions:

Reforming: $\quad a\ CH_3CH_2OH \rightarrow a\ CO + 3a\ H_2 + a\ C_{(ad)}$

Decarbonylation: $\quad b\ CH_3CH_2OH \rightarrow b\ CO + b\ CH_4 + b\ H_2$

Total decomposition: $\quad c\ CH_3CH_2OH \rightarrow 2c\ C_{(ad)} + c\ O_{(ad)} + 3c\ H_2$

The reforming reaction is desired to maximize hydrogen generation. The decarbonylation reaction produces undesired methane, and the total decomposition reaction leads to the accumulation of atomic carbon and oxygen and subsequent deactivation of the catalyst. In the above equations, the reforming activity is $a$, in units of monolayers as taken from the original reference, the total activity is $A = (a + b + c)$, and the relative selectivity for reforming is $S = a/A$. The reference data are reproduced in **Table S2** in the **Supporting Information**.

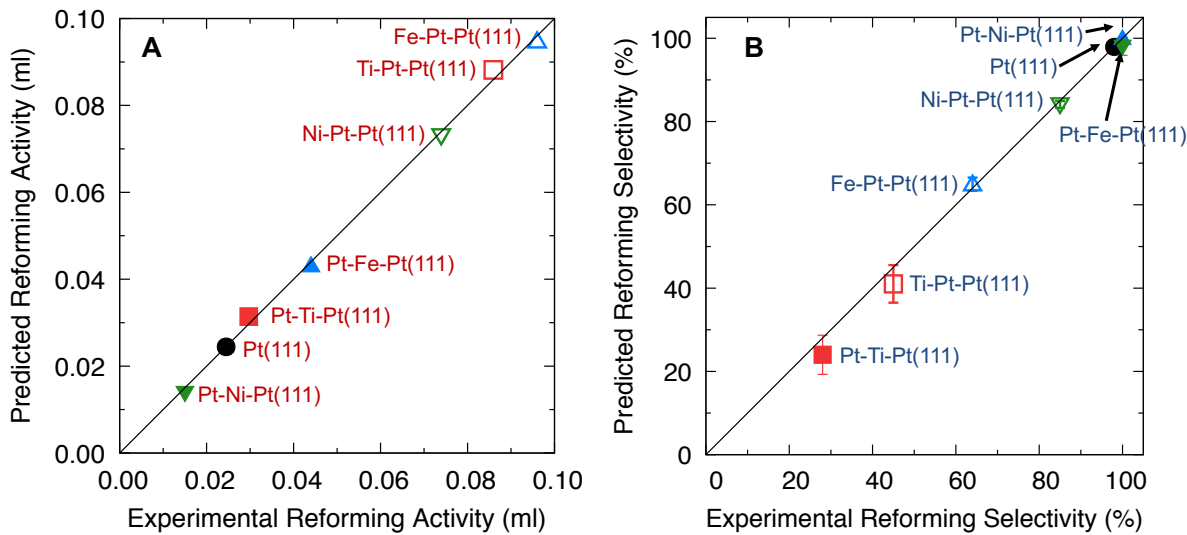

**Figure 4** Predictions by machine-learning Model 2 plotted against the experimental reference data for (**A**) ethanol reforming activity and (**B**) reforming selectivity. The plotted data points were obtained from leave-one-out cross validation. The RMSEs are 0.001 monolayers (ml) for the activity and 2.4% for the selectivity, respectively.

To construct predictive models of the reforming activity and selectivity based on only 7 reference data points, the model complexity has to be low, *i.e.*, the models have to be simple. The first step was therefore to identify those DFT reaction energies and transition-state energies that correlate most with the target quantities in order to minimize the number of required features. The reaction energies that enter ML Model 2 were obtained from additional DFT binding energy calculations, which are computationally more efficient and require less human intervention than transition-state energy calculations. We employed the Least-Absolute Shrinkage and Selection Operation (LASSO) method[41,42] to identify the most important reaction energies and transition-state energies, which were then used in subsequent non-regularized linear least-squares regression. The correlation of the resulting ML-predicted reforming activity and selectivity with the reference data is shown in **Figure 4**.

As seen in **Figure 4**, the linear ML models fit the reference data with good accuracy. The activity and selectivity models depend on four and five parameters, respectively, showing that the dataset of seven records is indeed sufficient. Note that the plots in **Figure 4** show the predictions obtained from leave-one-out cross validation, i.e., for each prediction only the data of 6 of the catalysts was used for the regression, and the 7th catalyst was predicted. This result confirms our hypothesis that the catalytic activity and selectivity are simple linear functions of the reaction energies and transition-state energies.

An advantage of the simple linear models is their interpretability. The reforming activity is determined by only two reactions, the C–C scission reaction (2) and the C–O scission reaction (8) of **Figure 2**:

$$*CH_3CHO \rightarrow *CH_3 + *CHO \quad (2)$$

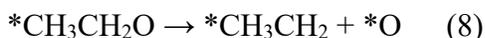

$$*CH_3CH_2O \rightarrow *CH_3CH_2 + *O \quad (8)$$

The model of the reforming activity $a$ as determined by linear regression analysis is (in units of monolayers)

$$a = 0.107\, E_{TS}^{(2)} - 0.128\, E_{TS}^{(8)} - 0.003\, E_r^{(8)} + 0.100 \ ,$$

where $E_{TS}^{(2)}$ and $E_{TS}^{(8)}$ are the transition-state energies and $E_r^{(8)}$ is the reaction energy of the C–O scission reaction (all energies in units of electronvolts). This means, the reforming activity $a$ increases as reaction (8) becomes more feasible (low transition-state and negative reaction energy). Additionally, the activity for reforming increases with increasing transition-state energy of the C–C scission reaction (2). The RMSE of the activity model is 0.001 monolayers as determined by leave-one-out cross validation (CV), and a three-parameter model without the reaction energy $E_r^{(8)}$ still results in a CV score of only 0.004 monolayers (**Figure S4**), showing that the linear model is robust. The accuracy and robustness of the model are remarkable given the uncertainty in the transition-state energies predicted by ML Model 1 and indicate that the relative error in the transition-state energy for reactions (2) and (8) across the different catalysts is smaller than the mean error of Model 1.

The reaction energies and transition-state energies of reactions (2) and (8) also correlate with the reforming selectivity, but the most robust selectivity model that we determined only depends on the three C–C scission reactions (2), (3), and (4), and does not involve the energetics from any C–O scission reaction. Since the selectivity $S$ can only vary between 0 and 1, the selectivity model was fitted to the *logit* of the experimental selectivity

$$S_l = \frac{1}{\beta} \ln \frac{S}{1-S}$$

and the selectivity is evaluated as the logistic function of the predicted logit selectivity $S_l$

$$S = \frac{1}{1 + e^{-\beta S_l}}$$

where the scaling parameter was set to $\beta = 10$ by empirical adjustment. The logit selectivity is given by

$$S_l = 1.705\, E_{TS}^{(2)} + 0.355\, E_r^{(2)} - 1.124\, E_{TS}^{(3)} - 1.274\, E_r^{(4)} - 0.487$$

and the three C–C scission reactions are

$$*CH_3CHO \rightarrow *CH_3 + *CHO \quad (2)$$
$$*CH_2CH_2O \rightarrow *CH_2 + *CH_2O \quad (3)$$
$$*CH_3CO \rightarrow *CH_3 + *CO \quad (4)$$

Thus, the reforming selectivity increases as reaction (2) becomes kinetically and thermodynamically less feasible, which is in agreement with the reforming activity model from above. On the other hand, fast kinetics (low transition-state energy) of reaction (3) and a thermodynamic driving force (low or negative reaction energy) for reaction (4) increase the reforming selectivity.

Since the selectivity model is more complex and still based on only 7 reference data points, we quantified the uncertainty of the model by constructing a third model that predicts the total activity $A$, from which the reforming selectivity can also be estimated as $S = a/A$ (see more details in **Section S4** of the **Supporting Information**). This second, independent model of the reforming selectivity is in good agreement with the logit selectivity model described above and was used to calculate the error bars in **Figure 4 B** and in the analysis described in the following.

Taken together, the activity and selectivity models give us some insight into the likely reaction mechanism of the ethanol reforming reaction. The strong correlation of both the reforming activity and selectivity with the transition-state energy of the C–C scission reaction (2) indicates that this reaction favors the competing ethanol decomposition pathways, such as methane production. The preference for reaction (4) is also in agreement with chemical intuition as it leads to the desired carbon monoxide production. The roles of reactions (3) and (8) are less obvious, and it is possible that the ethanol reforming mechanism varies among the catalysts.

It is also important to note that the reaction energies of the different C–C and C–O scission reactions and their transition-state energies are not fully independent. It has been well established that *scaling relations* exist that couple the adsorption energies of all oxophilic and all carbophilic intermediates on transition-metal surfaces.[43] It is therefore possible that the ethanol reforming reaction proceeds via C–C scission reactions other than reactions (3) and (4) depending on the catalyst.

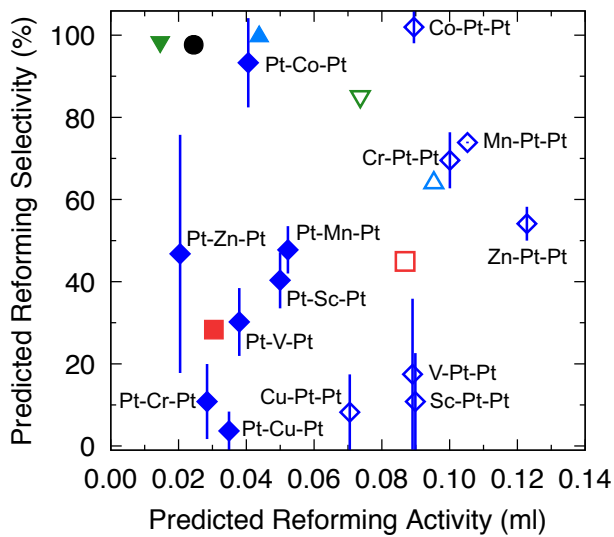

**Figure 5** Reforming activities and selectivities of core-shell catalysts. Blue diamond symbols indicate the predicted values for the compositions TM-Pt-Pt(111) (empty symbols) and Pt-TM-Pt(111) (filled symbols) with TM = Sc, V, Cr, Mn, Co, Cu, and Zn. The estimated uncertainty of the selectivity is shown as error bars. For comparison, the values of the experimentally characterized materials for TM = Ti, Fe, Ni, and Pt are also shown with the same colors and symbols as in **Figure 4**.

Beyond interpreting the available experimental data, the transition-state ML Model 1 and the activity/selectivity ML Model 2 allow, in combination, the prediction of reforming activities and selectivities based only on DFT reaction energies, which can be efficiently evaluated. We therefore calculated the DFT reaction energies for the hypothetical Pt-based core-shell architectures of the remaining 3d transition metals Sc, V, Cr, Mn, Co, and Zn. The 248 DFT energies of the reaction intermediates, which enter ML Model 1 as features, are provided as a comma-separated value file in the associated GitHub repository, along with their transition-state energies for the C–C and C–O scission reactions as predicted by ML Model 1. The uncertainty quantification of the predicted reforming selectivities is shown in **Figure S5**, and **Figure 5** shows the predicted reforming activities and selectivities.

As seen in **Figure 5**, the core-shell architectures terminated by a Pt monolayer have generally lower reforming activities than those terminated by the other transition metals. The selectivity varies widely with the transition-metal species, and uncertainty estimates are shown as error bars in the figure. Most of the considered catalyst compositions are predicted to exhibit either poorer reforming activity or poorer selectivity than those already experimentally characterized, but the catalytic properties of four of the hypothetical compositions are promising: Cr-Pt-Pt(111), Mn-Pt-Pt(111), Co-Pt-Pt(111), and Zn-Pt-Pt(111).

While this prediction is encouraging and warrants further experimental investigation of these materials, we note that our computational screening has not considered the synthesizability of the core-shell architectures. The computational prediction of the synthesizability is, in our experience, not trivial. For example, Ni forms a binary alloy with Pt, and DFT calculations predict the alloy to be more stable than the core-shell structure, even though both core-shell structures Ni-Pt-Pt(111) and Pt-Ni-Pt(111) can in practice be synthesized.[4] Hence, an experimental investigation of the predicted catalyst compositions will be needed to complete the assessment.

In conclusion, in this letter we demonstrated how a combination of machine learning and first-principles calculations can be used to extract knowledge from small experimental datasets both for the interpretation of experimental results and for the computational discovery of new catalysts. By combining a complex ML model trained on extensive computational data with simple linear regression models of experimental catalytic activities and selectivities from the literature, we could identify the key C–C bond scission reactions involved in the ethanol reforming reaction. The combined ML model based on computed and experimentally measured data was then used in a computational screening for ethanol reforming core-shell catalysts, which identified four promising compositions that have, to our knowledge, not yet been investigated.

## ASSOCIATED CONTENT

The Supporting Information is available free of charge at https://pubs.acs.org/doi/XXX. The reference data set containing the DFT binding and transition-state energies and the experimental activities and selectivities from the literature as well as the Python source code implementing all models can be obtained from the GitHub repository https://github.com/atomisticnet/ml-catalysis.

## ACKNOWLEDGMENTS

DFT calculations and machine-learning model construction made use of the Extreme Science and Engineering Discovery Environment (XSEDE), which is supported by National Science Foundation grant number ACI-1053575 (allocation no. DMR14005). Calculations were also performed on the computational resources of the Center for Functional Nanomaterials, which is a U.S. DOE Office of Science Facility, at Brookhaven National Laboratory under Contract No. DE-SC0012704. We also acknowledge computing resources from Columbia University's Shared Research Computing Facility project, which is supported by NIH Research Facility Improvement Grant 1G20RR030893-01, and associated funds from the New York State Empire State Development, Division of Science Technology and Innovation (NYSTAR) Contract C090171, both awarded April 15, 2010. This article was sponsored by the Catalysis Center for Energy Innovation (CCEI), an Energy Frontier Research Center (EFRC) funded by the U.S. Department of Energy, Office of Basic Energy Sciences under Award Number DE-SC0001004. N.A. thanks Dr. Jose Garrido Torres and Dr. Mark S Hybertsen for discussions.

**For Table of Contents Only**

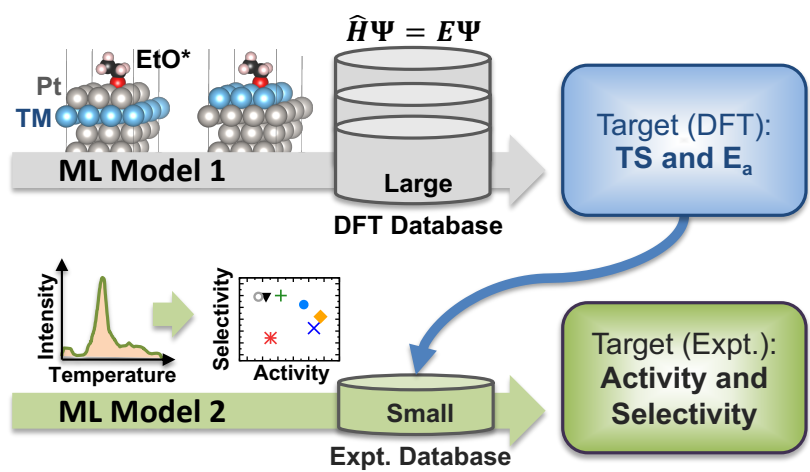

# Predicting the Activity and Selectivity of Bimetallic Metal Catalysts for Ethanol Reforming using Machine Learning

Nongnuch Artrith[1,*], Zhexi Lin[1], Jingguang G. Chen[1,2,*]

[1]*Department of Chemical Engineering, Columbia University, New York, NY, USA.*
[2]*Chemistry Division, Brookhaven National Laboratory, Upton, NY, USA.*
Corresponding authors: *nartrith@atomistic.net *jgchen@columbia.edu

## S1. Details of the density-functional theory (DFT) calculations

All DFT calculations were performed using the exchange-correlation functional by Perdew and Wang (PW91)[1] and projector-augmented wave (PAW) pseudopotentials[2] as implemented in the Vienna Ab-Initio Simulation Package (VASP).[3,4] A plane-wave basis set with cutoff of 400 eV was used for the representation of the Kohn-Sham orbitals, and a k-point mesh with a density of 25 Å$^{-1}$ was employed for the Brillouin-zone integration. The convergence criterion for the self-consistent energy was $10^{-5}$ eV, and in geometry optimizations the atomic forces were generally minimized until the residual forces were below 0.01 eV/Å. For the Pt(111) surface, adsorption energies calculated with spin-polarized DFT calculations were found to be within 5 meV of spin-averaged calculations, so that spin polarization was not considered for the reaction path calculations on any of the surfaces. VASP input files were generated using the *python materials genomic* (pymatgen) toolkit.[5]

Transition states were determined using the machine-learning accelerated nudged-elastic-band (NEB) method by Torres et al.[6–8] as implemented in the atomic simulation environment (ASE) package.[9] A convergence threshold of 0.05 eV/Å for interatomic forces was used for the optimization of the minimum energy pathway, and at least 15 NEB images were employed for the interpolation. The method by Smidstrup *et al.*[10] was used for the initial NEB interpolation.

## S2. Details of the structure models

All catalyst surfaces were modeled with the periodic surface slab approach with 3×3 surface unit cells. Asymmetric slab models with 5 transition-metal planes were constructed. The bottom two Pt layers were fixed on their ideal bulk positions, and the positions of the atoms within the top three layers were generally optimized. The width of the vacuum region was 15 Å.

The geometries of all reaction intermediates on the different catalyst surfaces were fully optimized. Different adsorption sites were tested for the Pt(111) surface, and the most stable geometry for Pt(111) was used as initial structure for the other transition-metal (111) surfaces.

## S3. Construction of the machine-learning (ML) models

All ML models were constructed using the scikit-learn software package.[11] Both the random-forest regression (RFR) model and the Gaussian process regression (GPR) model for transition-state energies were trained using the data in **Table S1**. Model features were

1. The atomic nearest-neighbor distances in the ideal crystal structures of the elements in the top monolayer, the second monolayer, and the bulk;
2. The Pauling electronegativities;
3. The reaction number in Figure 2 of the main manuscript;
4. A flag indicating whether the reaction is a C–C bond scission or a C–O bond scissions;
5. The DFT adsorption energy of ethanol on the surface; and
6. The DFT energies of the initial and final states of the reaction as well as the reaction energy.

Target for the training was the DFT transition-state energy. All energies are relative to the energy of the bare surface and the desorbed ethanol molecule, and the energy of the removed hydrogen atoms was included as multiples of the energy of an $H_2$ molecule. All features were standardized by removing the mean and scaling to unit variance using the standard scaler provided by scikit-learn.[11]

Both random forest regression (RFR) and Gaussian process regression (GPR) models rely on hyperparameters that need to be decided before the models are fitted. In the case of RFR, the hyperparameters are the maximal depth and the number of estimators, which were determined using an exhaustive grid search with 5-fold cross validation, as implemented by scikit-learn's 'GridSearchCV' class.[11]

In the case of GPR, the kernel parameters are hyperparameters. The kernel of our GPR model is the sum of an isotropic radial basis function (RBF) kernel, which depends on a length scale parameter, and a white noise kernel that depends on an amplitude. For the model fit, we made use of the scikit-learn class 'GaussianProcessRegressor' that optimizes the hyperparameters, i.e., the length scale of the RBF and the amplitude of the noise kernel, during the model fit.[11]

The RFR model had a maximal depth of 9 and was comprised of 15 estimators. The GPR model was based on a radial basis-function kernel with optimized length scale 2.11.

### S4. Selectivity model based on the total activity

The linear model for the total activity, as determined in the same fashion as the model for the reforming activity, is

$$A = 0.023\ E_{TS}^{(6)} - 0.113\ E_{TS}^{(8)} + 0.2080.113\ E_{r}^{(4)} - 0.045\ E_{r}^{(5)} + 0.265$$

which achieves a leave-one-out cross-validation score of 0.001 ML. In combination with the model for the reforming activity $a$, the reforming selectivity is given by $S = a/A$. Predictions by this linear selectivity model were compared with those of the logit selectivity discussed in the main manuscript to perform the uncertainty quantification shown in **Figure S5**.

## S5. Supporting figures

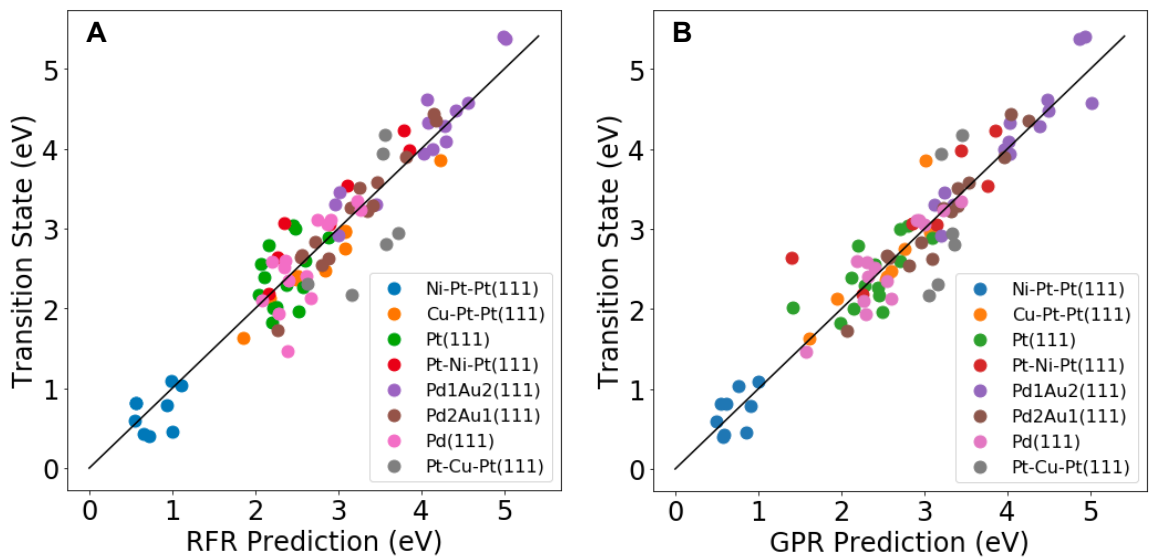

**Figure S1** Comparison of the transition-state energy ML models based on **(A)** random forest regression (RFR) and **(B)** Gaussian process regression (GPR). The plotted points were not included in the model construction and are predictions based on leave-one-out cross validation (CV).

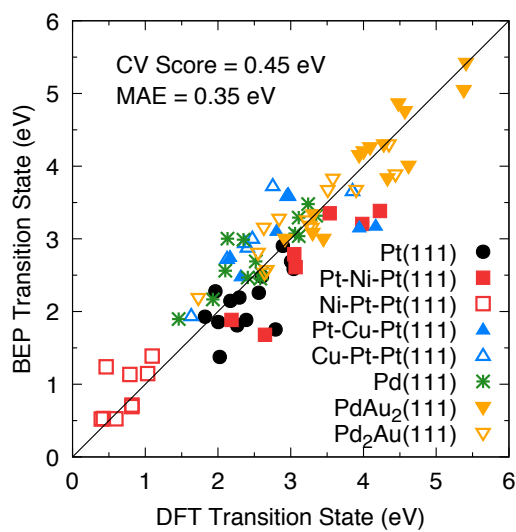

**Figure S2** Transition-state energies predicted with the Brønsted-Evans-Polanyi (BEP) principle[12,13] compared to the DFT reference energies. The plotted points were not included in the model construction and are predictions based on leave-one-out cross validation (CV).

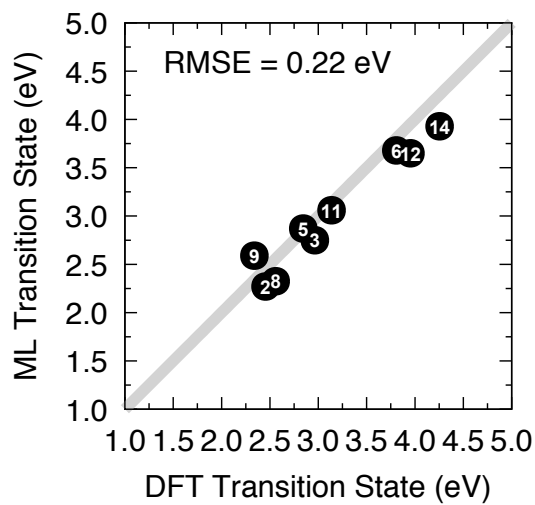

**Figure S3** Transition-state energies for different reactions over Pt-Fe-Pt(111) as predicted by ML Model 1 compared to the actual DFT energies. The root mean squared error (RMSE) of the ML predictions is 0.22 eV. The labels refer to the reaction numbers in **Figure 2** of the main manuscript.

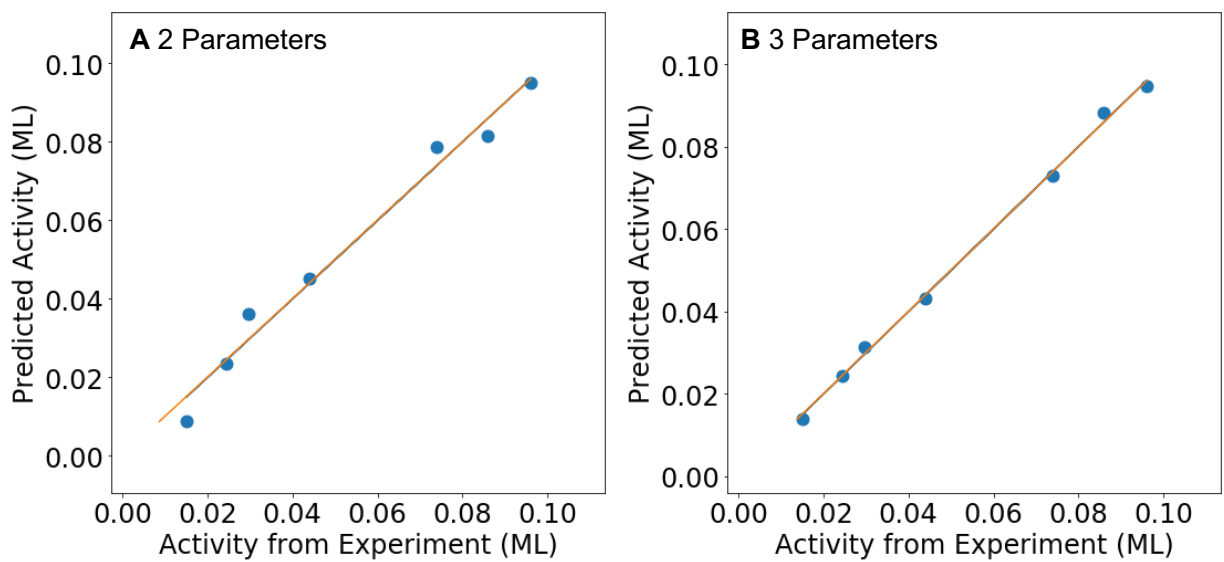

**Figure S4** Comparison of the activity models depending on **(A)** 2 parameters ($E_{TS}^{(2)}$, $E_{TS}^{(8)}$) and **(B)** 3 parameters ($E_{TS}^{(2)}$, $E_{TS}^{(8)}$ and $E_r^{(8)}$). The $y$ intercept is an additional model parameter.

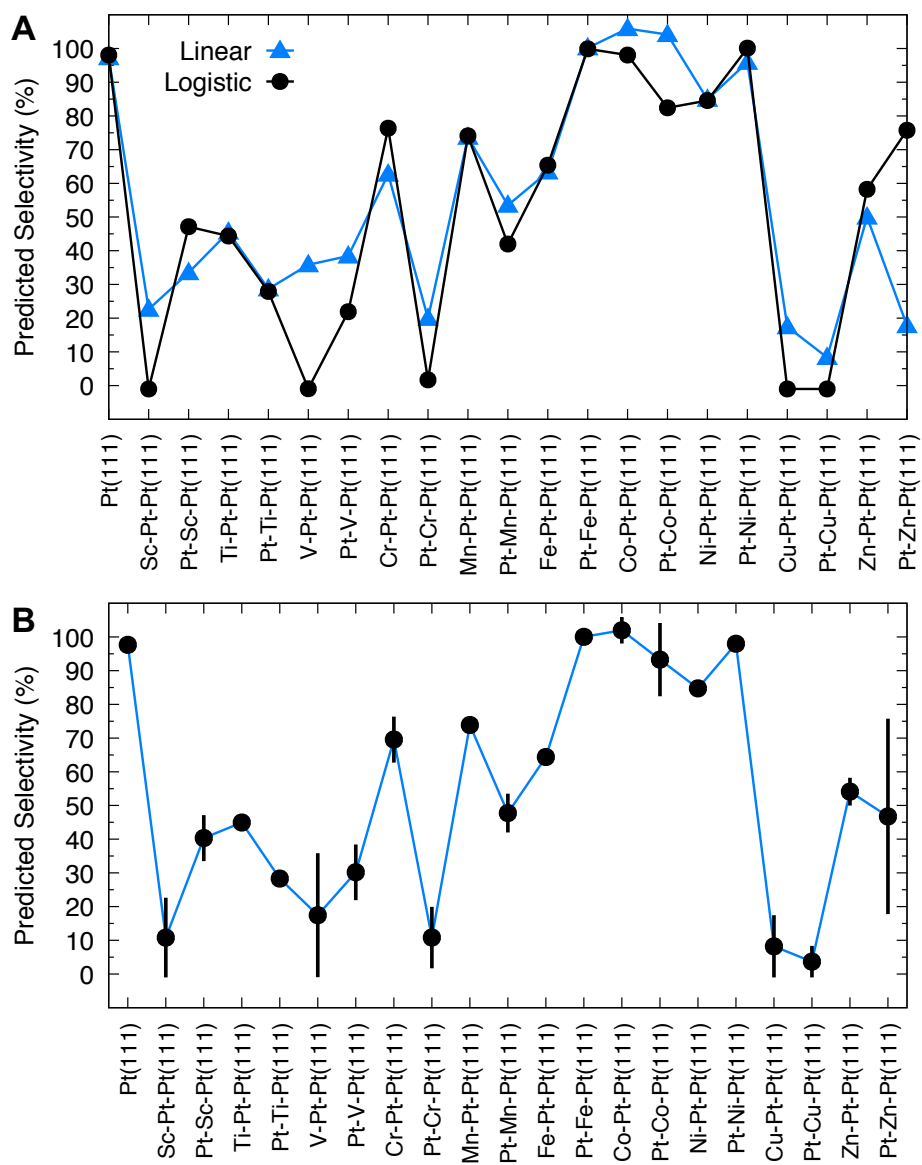

**Figure S5** Uncertainty quantification of the predicted selectivities. (**A**) Comparison of the predictions by the linear (blue triangles) and the logistic (black circles) selectivity models. (**B**) Representation of the predicted selectivities as the mean of the two models with error bars given by the difference of the model predictions.

## S6. Supporting Tables

**Table S1** Reference dataset used for the construction of the transition-state energy ML Model 1. $d_{NN}^1$, $d_{NN}^2$, and $d_{NN}^{bulk}$ are the atomic nearest-neighbor distances in the ideal crystal structures of the elements in the top monolayer, the second monolayer, and the bulk, respectively, taken from the book by Kittel.[14] $\chi^1$ and $\chi^2$ are the Pauling electronegativities.[15] ID is the reaction number in Figure 2 of the main manuscript. CO/CC is equal to 1 for C–C bond scission reactions and equal to 2 for C–O bond scissions. $E_{ads}^{EtOH}$ is the DFT adsorption energy of ethanol on the surface. $E_{is}$, $E_{fs}$, and $E_r$ are the DFT energies of the initial and final states of the reaction as well as the reaction energy. $E_{TS}$ is the transition-state energy. All energies are relative to the energy of the bare surface and the desorbed ethanol molecule. Energies for Pd the PdAu alloys were taken from Li et al.[16]

| Surface | $d_{NN}^1$ | $d_{NN}^2$ | $d_{NN}^{bulk}$ | $\chi^1$ | $\chi^2$ | ID | CO/CC | $E_{ads}^{EtOH}$ | $N_H$ | $E_{is}$ | $E_{fs}$ | $E_r$ | $E_{TS}$ |
|---|---|---|---|---|---|---|---|---|---|---|---|---|---|
| | Å | Å | Å | | | | | eV | | eV | eV | eV | eV |
| Cu-Pt-Pt(111) | 256 | 277 | 277 | 1.90 | 2.28 | 4 | 1 | -0.49570 | 3 | 0.7059 | 2.3669 | 1.6610 | 2.3669 |
| Cu-Pt-Pt(111) | 256 | 277 | 277 | 1.90 | 2.28 | 5 | 1 | -0.49570 | 3 | 0.5158 | 2.4030 | 1.8872 | 2.4030 |
| Cu-Pt-Pt(111) | 256 | 277 | 277 | 1.90 | 2.28 | 6 | 1 | -0.49570 | 3 | 2.0088 | 2.7528 | 0.7440 | 2.7528 |
| Cu-Pt-Pt(111) | 256 | 277 | 277 | 1.90 | 2.28 | 7 | 1 | -0.49570 | 2 | 1.4318 | 2.9658 | 1.5340 | 2.9658 |
| Cu-Pt-Pt(111) | 256 | 277 | 277 | 1.90 | 2.28 | 8 | 2 | -0.49570 | 5 | -0.6443 | 1.6318 | 2.2761 | 1.6317 |
| Cu-Pt-Pt(111) | 256 | 277 | 277 | 1.90 | 2.28 | 9 | 2 | -0.49570 | 4 | 0.5202 | 2.1252 | 1.6050 | 2.1252 |
| Cu-Pt-Pt(111) | 256 | 277 | 277 | 1.90 | 2.28 | 11 | 2 | -0.49570 | 3 | 0.7059 | 2.4719 | 1.7660 | 2.4719 |
| Cu-Pt-Pt(111) | 256 | 277 | 277 | 1.90 | 2.28 | 12 | 2 | -0.49570 | 3 | 0.5158 | 3.8498 | 3.3340 | 3.8498 |
| Cu-Pt-Pt(111) | 256 | 277 | 277 | 1.90 | 2.28 | 14 | 2 | -0.49570 | 2 | 1.4318 | 2.9568 | 1.5250 | 2.9568 |
| Ni-Pt-Pt(111) | 249 | 277 | 277 | 1.91 | 2.28 | 2 | 1 | -0.73068 | 4 | -0.5303 | -0.9524 | -0.4221 | 0.4237 |
| Ni-Pt-Pt(111) | 249 | 277 | 277 | 1.91 | 2.28 | 4 | 1 | -0.73068 | 3 | -0.2382 | -0.8449 | -0.6067 | 0.8172 |
| Ni-Pt-Pt(111) | 249 | 277 | 277 | 1.91 | 2.28 | 5 | 1 | -0.73068 | 3 | -0.1366 | -0.1891 | -0.0525 | 0.7884 |
| Ni-Pt-Pt(111) | 249 | 277 | 277 | 1.91 | 2.28 | 6 | 1 | -0.73068 | 3 | 0.2829 | -0.3591 | -0.6419 | 0.4621 |
| Ni-Pt-Pt(111) | 249 | 277 | 277 | 1.91 | 2.28 | 7 | 1 | -0.73068 | 2 | 0.5864 | -0.3131 | -0.8994 | 1.0954 |
| Ni-Pt-Pt(111) | 249 | 277 | 277 | 1.91 | 2.28 | 9 | 2 | -0.73068 | 4 | -0.5303 | -0.9558 | -0.4255 | 0.5965 |
| Ni-Pt-Pt(111) | 249 | 277 | 277 | 1.91 | 2.28 | 11 | 2 | -0.73068 | 3 | -0.2382 | -1.2013 | -0.9631 | 0.3952 |
| Ni-Pt-Pt(111) | 249 | 277 | 277 | 1.91 | 2.28 | 12 | 2 | -0.73068 | 3 | -0.1366 | -0.9578 | -0.8212 | 0.8154 |
| Ni-Pt-Pt(111) | 249 | 277 | 277 | 1.91 | 2.28 | 14 | 2 | -0.73068 | 2 | 0.5864 | -0.7332 | -1.3196 | 1.0321 |
| Pd(111) | 275 | 275 | 275 | 2.20 | 2.20 | 1 | 1 | -0.73 | 5 | 0.6500 | 1.56 | 0.91 | 2.5800 |
| Pd(111) | 275 | 275 | 275 | 2.20 | 2.20 | 2 | 1 | -0.73 | 4 | 1.0800 | 1.34 | 0.26 | 2.6000 |
| Pd(111) | 275 | 275 | 275 | 2.20 | 2.20 | 3 | 1 | -0.73 | 4 | 1.6400 | 1.72 | 0.08 | 2.3500 |
| Pd(111) | 275 | 275 | 275 | 2.20 | 2.20 | 4 | 1 | -0.73 | 3 | 0.9600 | 0.3 | -0.66 | 1.4600 |

| | | | | | | | | | | | | |
|---|---|---|---|---|---|---|---|---|---|---|---|---|
| Pd(111) | 275 | 275 | 275 | 2.20 | 2.20 | 5 | 1 | -0.73 | 3 | 1.4300 | 2.05 | 0.62 | 3.0600 |
| Pd(111) | 275 | 275 | 275 | 2.20 | 2.20 | 6 | 1 | -0.73 | 3 | 2.1900 | 1.84 | -0.35 | 3.1100 |
| Pd(111) | 275 | 275 | 275 | 2.20 | 2.20 | 7 | 1 | -0.73 | 2 | 1.5800 | 1 | -0.58 | 2.1000 |
| Pd(111) | 275 | 275 | 275 | 2.20 | 2.20 | 8 | 2 | -0.73 | 5 | 0.6500 | 1.04 | 0.39 | 1.9300 |
| Pd(111) | 275 | 275 | 275 | 2.20 | 2.20 | 9 | 2 | -0.73 | 4 | 1.0900 | 1.63 | 0.54 | 2.5200 |
| Pd(111) | 275 | 275 | 275 | 2.20 | 2.20 | 10 | 2 | -0.73 | 4 | 1.6400 | 1.74 | 0.1 | 2.1300 |
| Pd(111) | 275 | 275 | 275 | 2.20 | 2.20 | 11 | 2 | -0.73 | 3 | 0.9600 | 1.34 | 0.38 | 2.4100 |
| Pd(111) | 275 | 275 | 275 | 2.20 | 2.20 | 12 | 2 | -0.73 | 3 | 1.4400 | 1.98 | 0.54 | 3.1100 |
| Pd(111) | 275 | 275 | 275 | 2.20 | 2.20 | 13 | 2 | -0.73 | 3 | 2.1800 | 2.18 | 0 | 3.2400 |
| Pd(111) | 275 | 275 | 275 | 2.20 | 2.20 | 14 | 2 | -0.73 | 2 | 1.5800 | 2.37 | 0.79 | 3.3500 |
| Pd1Au2(111) | 284 | 284 | 284 | 2.43 | 2.43 | 1 | 1 | -0.62 | 5 | 1.2900 | 2.1700 | 0.8800 | 3.3000 |
| Pd1Au2(111) | 284 | 284 | 284 | 2.43 | 2.43 | 2 | 1 | -0.62 | 4 | 1.0600 | 2.2100 | 1.1500 | 3.4500 |
| Pd1Au2(111) | 284 | 284 | 284 | 2.43 | 2.43 | 3 | 1 | -0.62 | 4 | 2.4300 | 3.3700 | 0.9400 | 4.0900 |
| Pd1Au2(111) | 284 | 284 | 284 | 2.43 | 2.43 | 4 | 1 | -0.62 | 3 | 1.2500 | 2.0500 | 0.8000 | 2.9200 |
| Pd1Au2(111) | 284 | 284 | 284 | 2.43 | 2.43 | 5 | 1 | -0.62 | 3 | 2.4600 | 3.4200 | 0.960 | 4.2800 |
| Pd1Au2(111) | 284 | 284 | 284 | 2.43 | 2.43 | 6 | 1 | -0.62 | 3 | 3.3900 | 3.690 | 0.300 | 4.4800 |
| Pd1Au2(111) | 284 | 284 | 284 | 2.43 | 2.43 | 7 | 1 | -0.62 | 2 | 2.4000 | 3.200 | 0.800 | 3.9400 |
| Pd1Au2(111) | 284 | 284 | 284 | 2.43 | 2.43 | 8 | 2 | -0.62 | 5 | 1.2900 | 2.6200 | 1.3300 | 3.3100 |
| Pd1Au2(111) | 284 | 284 | 284 | 2.43 | 2.43 | 9 | 2 | -0.62 | 4 | 1.0700 | 3.7300 | 2.6600 | 4.3300 |
| Pd1Au2(111) | 284 | 284 | 284 | 2.43 | 2.43 | 10 | 2 | -0.62 | 4 | 2.4300 | 3.2700 | 0.8400 | 4.0000 |
| Pd1Au2(111) | 284 | 284 | 284 | 2.43 | 2.43 | 11 | 2 | -0.62 | 3 | 1.2500 | 3.9100 | 2.6600 | 4.6200 |
| Pd1Au2(111) | 284 | 284 | 284 | 2.43 | 2.43 | 12 | 2 | -0.62 | 3 | 2.4700 | 4.2200 | 1.7500 | 4.5700 |
| Pd1Au2(111) | 284 | 284 | 284 | 2.43 | 2.43 | 13 | 2 | -0.62 | 3 | 3.3900 | 4.690 | 1.300 | 5.4100 |
| Pd1Au2(111) | 284 | 284 | 284 | 2.43 | 2.43 | 14 | 2 | -0.62 | 2 | 2.4000 | 4.830 | 2.430 | 5.3800 |
| Pd2Au1(111) | 279 | 279 | 279 | 2.31 | 2.31 | 1 | 1 | -0.83 | 5 | 0.9600 | 1.45 | 0.49 | 2.6400 |
| Pd2Au1(111) | 279 | 279 | 279 | 2.31 | 2.31 | 2 | 1 | -0.83 | 4 | 0.9000 | 1.56 | 0.66 | 2.6700 |
| Pd2Au1(111) | 279 | 279 | 279 | 2.31 | 2.31 | 3 | 1 | -0.83 | 4 | 1.8700 | 2.76 | 0.89 | 3.5100 |
| Pd2Au1(111) | 279 | 279 | 279 | 2.31 | 2.31 | 4 | 1 | -0.83 | 3 | 0.8900 | 0.87 | -0.02 | 1.7300 |
| Pd2Au1(111) | 279 | 279 | 279 | 2.31 | 2.31 | 5 | 1 | -0.83 | 3 | 1.4200 | 2.37 | 0.95 | 3.2200 |
| Pd2Au1(111) | 279 | 279 | 279 | 2.31 | 2.31 | 6 | 1 | -0.83 | 3 | 2.5900 | 2.47 | -0.12 | 3.5800 |
| Pd2Au1(111) | 279 | 279 | 279 | 2.31 | 2.31 | 7 | 1 | -0.83 | 2 | 1.6400 | 2.01 | 0.37 | 2.6300 |
| Pd2Au1(111) | 279 | 279 | 279 | 2.31 | 2.31 | 8 | 2 | -0.83 | 5 | 0.9600 | 1.93 | 0.97 | 2.5500 |
| Pd2Au1(111) | 279 | 279 | 279 | 2.31 | 2.31 | 9 | 2 | -0.83 | 4 | 0.9000 | 2.65 | 1.75 | 3.2600 |
| Pd2Au1(111) | 279 | 279 | 279 | 2.31 | 2.31 | 10 | 2 | -0.83 | 4 | 1.8700 | 2.05 | 0.18 | 2.8400 |
| Pd2Au1(111) | 279 | 279 | 279 | 2.31 | 2.31 | 11 | 2 | -0.83 | 3 | 0.8900 | 2.61 | 1.72 | 3.2900 |
| Pd2Au1(111) | 279 | 279 | 279 | 2.31 | 2.31 | 12 | 2 | -0.83 | 3 | 1.4300 | 3.12 | 1.69 | 3.9000 |
| Pd2Au1(111) | 279 | 279 | 279 | 2.31 | 2.31 | 13 | 2 | -0.83 | 3 | 2.5900 | 3.32 | 0.73 | 4.3500 |
| Pd2Au1(111) | 279 | 279 | 279 | 2.31 | 2.31 | 14 | 2 | -0.83 | 2 | 1.6400 | 3.35 | 1.71 | 4.4400 |
| Pt-Cu-Pt(111) | 277 | 256 | 277 | 2.28 | 1.90 | 3 | 1 | -0.24323 | 4 | 1.4395 | 2.9425 | 1.5030 | 2.9425 |
| Pt-Cu-Pt(111) | 277 | 256 | 277 | 2.28 | 1.90 | 4 | 1 | -0.24323 | 3 | 0.5163 | 2.1663 | 1.6500 | 2.1663 |
| Pt-Cu-Pt(111) | 277 | 256 | 277 | 2.28 | 1.90 | 7 | 1 | -0.24323 | 2 | 1.0929 | 1.2619 | 0.1690 | 2.3139 |
| Pt-Cu-Pt(111) | 277 | 256 | 277 | 2.28 | 1.90 | 11 | 2 | -0.24323 | 3 | 0.5163 | 2.8053 | 2.2890 | 2.8053 |

| Surface | | | | | | | | | | | | |
|---|---|---|---|---|---|---|---|---|---|---|---|---|
| Pt-Cu-Pt(111) | 277 | 256 | 277 | 2.28 | 1.90 | 12 | 2 | -0.24323 | 3 | 1.3624 | 2.2673 | 0.9049 3.9424 |
| Pt-Cu-Pt(111) | 277 | 256 | 277 | 2.28 | 1.90 | 14 | 2 | -0.24323 | 2 | 1.0929 | 2.5397 | 1.4468 4.1689 |
| Pt-Ni-Pt(111) | 277 | 249 | 277 | 2.28 | 1.91 | 4  | 1 | -0.28966 | 3 | 0.6273 | 0.5693 | -0.0580 2.1853 |
| Pt-Ni-Pt(111) | 277 | 249 | 277 | 2.28 | 1.91 | 5  | 1 | -0.28966 | 3 | 1.2521 | 1.6868 | 0.4347 3.0481 |
| Pt-Ni-Pt(111) | 277 | 249 | 277 | 2.28 | 1.91 | 6  | 1 | -0.28966 | 3 | 2.0580 | 2.0611 | 0.0031 3.5360 |
| Pt-Ni-Pt(111) | 277 | 249 | 277 | 2.28 | 1.91 | 8  | 2 | -0.28966 | 5 | 0.7582 | 0.1434 | -0.6148 2.6455 |
| Pt-Ni-Pt(111) | 277 | 249 | 277 | 2.28 | 1.91 | 11 | 2 | -0.28966 | 3 | 0.6273 | 1.8668 | 1.2395 3.0678 |
| Pt-Ni-Pt(111) | 277 | 249 | 277 | 2.28 | 1.91 | 12 | 2 | -0.28966 | 3 | 1.2521 | 2.4435 | 1.1914 3.9831 |
| Pt-Ni-Pt(111) | 277 | 249 | 277 | 2.28 | 1.91 | 14 | 2 | -0.28966 | 2 | 1.3187 | 2.7178 | 1.3991 4.2267 |
| Pt(111) | 277 | 277 | 277 | 2.28 | 2.28 | 1  | 1 | -0.37732 | 5 | 0.5771 | 0.3991 | -0.1780 2.7909 |
| Pt(111) | 277 | 277 | 277 | 2.28 | 2.28 | 2  | 1 | -0.37732 | 4 | 0.6068 | 0.5283 | -0.0784 2.0038 |
| Pt(111) | 277 | 277 | 277 | 2.28 | 2.28 | 3  | 1 | -0.37732 | 4 | 0.9086 | 1.0147 | 0.1061 2.5607 |
| Pt(111) | 277 | 277 | 277 | 2.28 | 2.28 | 4  | 1 | -0.37732 | 3 | 0.2367 | -0.0195 | -0.2562 2.0240 |
| Pt(111) | 277 | 277 | 277 | 2.28 | 2.28 | 5  | 1 | -0.37732 | 3 | 0.9065 | 0.8928 | -0.0137 2.2932 |
| Pt(111) | 277 | 277 | 277 | 2.28 | 2.28 | 6  | 1 | -0.37732 | 3 | 1.4296 | 1.3776 | -0.0520 3.0045 |
| Pt(111) | 277 | 277 | 277 | 2.28 | 2.28 | 7  | 1 | -0.37732 | 2 | 0.7565 | 0.5358 | -0.2206 1.8205 |
| Pt(111) | 277 | 277 | 277 | 2.28 | 2.28 | 8  | 2 | -0.37732 | 5 | 0.5771 | 0.4793 | -0.0978 2.2635 |
| Pt(111) | 277 | 277 | 277 | 2.28 | 2.28 | 9  | 2 | -0.37732 | 4 | 0.6068 | 1.0438 | 0.4370 2.1708 |
| Pt(111) | 277 | 277 | 277 | 2.28 | 2.28 | 10 | 2 | -0.37732 | 4 | 0.9086 | 1.0343 | 0.1257 1.9649 |
| Pt(111) | 277 | 277 | 277 | 2.28 | 2.28 | 11 | 2 | -0.37732 | 3 | 0.2367 | 0.8748 | 0.6381 2.3886 |
| Pt(111) | 277 | 277 | 277 | 2.28 | 2.28 | 12 | 2 | -0.37732 | 3 | 0.9065 | 1.4083 | 0.5018 2.6037 |
| Pt(111) | 277 | 277 | 277 | 2.28 | 2.28 | 13 | 2 | -0.37732 | 3 | 1.4296 | 1.7480 | 0.3184 2.8911 |
| Pt(111) | 277 | 277 | 277 | 2.28 | 2.28 | 14 | 2 | -0.37732 | 2 | 0.7565 | 1.7179 | 0.9614 3.0400 |

**Table S2** Activity and selectivity data by Skoplyak *et al.*[17,18] The reforming activity is the product of the total activity and the reforming selectivity.

| Surface | Total Activity (ML) | Reforming Selectivity | Decomp. Selectivity | Decarbonyl. Selectivity | Reference |
|---|---|---|---|---|---|
| Pt(111) | 0.025 | 0.980 | 0.000 | 0.020 | 17 |
| Pt-Ni-Pt(111) | 0.015 | 1.000 | 0.000 | 0.000 | 17 |
| Ni-Pt-Pt(111) | 0.087 | 0.850 | 0.030 | 0.120 | 17 |
| Pt-Fe-Pt(111) | 0.044 | 1.000 | 0.000 | 0.000 | 18 |
| Fe-Pt-Pt(111) | 0.150 | 0.640 | 0.360 | 0.000 | 18 |
| Pt-Ti-Pt(111) | 0.106 | 0.280 | 0.720 | 0.000 | 18 |
| Ti-Pt-Pt(111) | 0.191 | 0.450 | 0.550 | 0.000 | 18 |